\documentclass{aa}
\usepackage{graphics}

\begin{document}

   \title{Turbulent viscosity in clumpy accretion disks\\
	Application to the Galaxy}

   \titlerunning{Turbulent viscosity in clumpy accretion disks}

   \author{B.~Vollmer \and  T.~Beckert}

   \offprints{B.~Vollmer, e-mail: bvollmer@mpifr-bonn.mpg.de}

   \institute{Max-Planck-Institut f\"ur Radioastronomie, Auf dem
              H\"ugel 69, 53121 Bonn, Germany.}

   \date{Received / Accepted}

\abstract{The equilibrium state of a turbulent clumpy gas disk is analytically
investigated. The disk consists of distinct self-gravitating clouds.
Gravitational cloud--cloud interactions transfer energy over spatial scales
and produce a viscosity, which allows mass accretion in the gas disk.
Turbulence is assumed to be generated by instabilities 
involving self-gravitation and to be maintained by the energy input from
differential rotation and mass transfer. Disk parameters, global 
filling factors, molecular fractions, and star formation rates are derived.
The application of our model to the Galaxy shows good agreement with 
observations. They are consistent with the scenario where turbulence 
generated and maintained by gravitation 
can account for the viscosity in the gas disk of spiral galaxies.
The r\^{o}le of the galaxy mass for the morphological classification 
of spiral galaxies is investigated. 
 }

\maketitle

\section{Introduction}

Galactic gas disks are not continuous but clumpy. The structure
of the interstellar medium (ISM) is usually hierarchical (Scalo 1985)
over length scales of several magnitudes up to $\sim$100~pc. 
The clouds are not uniform
nor isolated and their boundaries are often of fractal nature
(Elmegreen \& Falgarone 1996). Whereas the atomic gas (H{\sc i})
is mainly in the form of filaments, the molecular gas is highly clumped.
The largest self-gravitating molecular clouds (giant molecular clouds = GMC)
have sizes of $\sim$30~pc and masses of $\sim 10^{5}{\mathrm M}_{\odot}$
(see e.g. Larson 1981). The GMCs have a volume filling factor
$\phi_{\rm V} \sim 10^{-4}$, with the ratio of the diameter of a
typical GMC to the vertical scale height of the GMC distribution
(=130~pc, Sanders et al. 1985) $\sim$0.4. In this respect, the GMC 
distribution resembles more a planetary ring (Goldreich \& Tremaine 1987).
The mass--size relation of the GMCs is $M \propto L^{D}$
The fractal dimension of a volume fractal is $D \sim 2.3$ (Elmegreen \&
Falgarone 1996). The origin of this dimension could be turbulent diffusion
in an incompressible fluid with a Kolmogorov velocity spectrum
($D \sim 2+\xi$ for $\Delta v \propto L^{\xi}$; see Meneveau \& Sreenivasan
1990). 

The Kolmogorov theory applies for fully developed subsonic incompressible
fluids. However, the ISM is supersonic and compressible.
Only recently, 3D numerical studies of magneto-hydrodynamical and 
hydrodynamical turbulence in an isothermal, compressible, and
self-gravitating fluid indicated that the energy spectrum of supersonic 
compressible turbulence follows a Kolmogorov law (Mac Low 1999, 
Klessen et al. 2000, Mac Low \& Ossenkopf 2000). The molecular clouds
themselves are stabilized against gravitational collapse by the turbulent
velocity field within them (see e.g. Larson 1981). 

Wada \& Norman (1999) used  high-resolution, 2D, hydrodynamical 
simulations to investigate the evolution of a self-gravitating multiphase 
interstellar medium in a galactic disk. They found that a gravitationally 
and thermally unstable disk evolves towards a globally quasi-stationary state
where the disk is characterized by clumpy and filamentary structures.
The energy source of the turbulence in this system originate in the shear 
driven by galactic rotation and self-gravitational energy of the gas. 
The effective Q parameter of the disk was found to have a value between 
2 and 5. Without feedback the energy spectrum $E(k) \propto k^{-3}$ corresponds
to a Kolmogorov law in two dimensions, but changes into $E(k) \propto k^{-2}$
if stellar energy feedback is included (Wada \& Norman 2001). 
This power law is expected if shocks dominate the system (Passot et al. 1988). 
Furthermore, Wada \& Norman (2001) derived a driving length scale of $\sim$200~pc for the model without stellar energy feedback.

While the turbulent nature of the ISM is well established now,
its origin and maintenance is still a matter of debate. 
This is of great importance, because turbulence can provide
angular momentum transport in the disk of spiral galaxies.
The evolution and the structure of an accretion disk depends
entirely on the effective viscosity caused by turbulence.

Two ingredients are necessary for a steady state turbulence:
(i) a dynamical instability to generate and (ii) a steady energy input
to maintain turbulence.\\
(i) Five possible instabilities were put forward by different authors:
\begin{itemize}
\item
viscous instability (Lightman \& Eardley 1974)
\item
thermal instability (Pringle et al. 1973)
\item
Parker instability (Parker 1966)
\item
magneto-hydrodynamic instability (Balbus \& Hawley 1991)
\item
gravitational instability (Toomre 1964).
\end{itemize}
Because self-gravity is always present in the 
ISM, the gravitational instability can operate in conjunction with
all other forms of instabilities. As soon as the density rises
due to such a combined instability and increases over the value
of the ambient medium by a factor 2, self-gravitation should be 
the dominant force (Elmegreen 1982).
Moreover, the gravitational instability can produce 
a whole hierarchy of clumped structures inside the largest ones.
Therefore, we will only discuss gravitational instability.

(ii) Two main mechanisms are proposed to provide the energy input in 
order to maintain the observed turbulence of 
$v_{\rm turb} \sim 10$~km\,s$^{-1}$:
\begin{itemize}
\item
supernova explosions (see e.g. Spitzer 1968, McKee \& Ostriker 1977)
\item
galactic differential rotation (Fleck 1981).
\end{itemize}
The supernova shocks, which can accelerate low mass clouds,
are extremely ineffective in accelerating GMCs, because of the
much larger mass to area ratio of GMCs. Jog \& Ostriker (1988)
applying directly the results of McKee \& Ostriker (1977) found
that supernova shocks can provide no more than 10\% of the
kinetic energy of the GMCs (however, Cui et al. (1996) suggested that
uncertain background corrections could make this fraction larger). 
Therefore, we will only focus on energy input by galactic rotation.

The effective viscosity $\nu_{\rm eff}$ due to turbulence can be
expressed in general as
\begin{equation}
\nu_{\rm eff}=\xi\,v_{\rm turb}\,l_{\rm driv}\ , 
\end{equation}
where $v_{\rm turb}$ is the turbulent velocity on large scales, 
$l_{\rm driv}$ the driving wavelength, and $\xi$ a constant.
For continuous disks the most widely used form is the so-called 
$\alpha$--ansatz introduced by Shakura (1972) and Shakura \& Sunyaev (1973). 
They assumed that the turbulent velocity is limited by the sound velocity 
($v_{\rm turb}=c_{\rm s}$),
the turbulent length scale equals the disk scale height $l_{\rm driv}=H$, 
and $\xi=\alpha < 1$. Lin \& Pringle (1987a) derived a prescription for the 
effective viscosity due to gravitational instability:
$l_{\rm driv}=H$ and $\xi=Q^{-2}$, where $Q$ is the Toomre parameter
(Toomre 1964). Recently, Duschl et al. (2000) suggested a
prescription for the viscosity due to hydrodynamically driven turbulence
at the critical effective Reynolds number $Re_{\rm crit}$, where
$\xi=1$, $v_{\rm turb}=Re_{\rm crit}^{-\frac{1}{2}}\,v_{\rm rot}$, and 
the driving wavelength is a well defined fraction of the local radius
in the disk $l_{\rm driv}=Re_{\rm crit}^{-\frac{1}{2}}\,R$. 

For clumpy accretion disks, Goldreich \& Tremaine (1978)
and Stewart \& Kaula (1980) elaborated models where the shear viscosity
in a rotating disk is due to cloud--cloud interactions. Their prescription
has the form:
\begin{equation}\label{eq:goldreich}
\nu_{\rm eff}=\frac{v_{\rm turb}^{2}}{\Omega}\frac{\tau}{\tau^{2}+1}\simeq
\tau^{-1}\frac{v_{\rm turb}^{2}}{\Omega}\ \ {\rm for}\ \tau\gg 1,
\end{equation}
where $\Omega$ is the rotational angular velocity and 
$\tau=\Omega\,t_{\rm int}$ is the number of cloud interactions per rotation 
period with the interaction time scale $t_{\rm int}$ between the clouds.
Vertical hydrostatic equilibrium $R/H \sim v_{\rm rot}/v_{\rm turb}$ gives
\begin{equation}
\nu_{\rm eff}=\tau^{-1}v_{\rm turb}H\ ,
\label{eq:gtsk}
\end{equation}
where $v_{\rm rot}=\Omega\,R$ is the rotation velocity.
In this article we derive a prescription for the turbulent
viscosity of comparable form with $\xi=Re^{-1}$ in terms of the turbulent 
Reynolds number $Re$. All disk parameters are given as functions of
$Re$, $Q$, and the mass accretion rate $\dot{M}$. We apply our
model to the Galaxy and derive a Galactic mass accretion rate.

\section{The basic picture}

We consider a gaseous accretion disk in a given gravitational
potential $\Phi$ which gives rise to an angular velocity
$\Omega=\sqrt{R^{-1}\frac{{\rm d}\Phi}{{\rm d}R}}$.
The Toomre parameter (Toomre 1964) is treated as a free parameter
\begin{equation}
Q=\frac{v_{\rm turb}\,\kappa}{\pi\,G\,\Sigma}\ ,
\end{equation}
 with the restriction $Q \geq 1$, where $G$ is the gravitational constant,
$\Sigma$ the average gas column density, and $\kappa$ the local
epicyclic frequency.
The disk consists of distinct self-gravitating clouds, which are
orbiting in an external gravitational field and have a velocity
dispersion $\Delta v$. These clouds might be embedded in a low density medium 
as long as $\Delta v/c_{\rm s}^{\rm ext} \leq 1$, where 
$c_{\rm s}^{\rm ext}$ is the sound speed of the external medium.
This is the case for the warm atomic or ionized phase ($T\sim 8000$~K)
and the hot ionized phase ($T\sim 10^{6}$~K).
The disk scale height is determined uniquely by the turbulent 
pressure $p_{\rm turb}=\rho v_{\rm turb}^{2}$, where $\rho$
is the averaged density in the disk.
The clouds have a small volume filling factor $\phi_{\rm V} \ll 1$.
The transport of angular momentum is due to cloud--cloud 
interactions. These can be gravitational (elastic) encounters or
direct (inelastic) collisions. Thus, gravitational scattering of the
massive clouds off each other in a differentially rotating galactic disk
gives rise to an effective ``gravitational'' viscosity. 
Jog \& Ostriker (1988) determined analytically the one-dimensional
velocity dispersion of such a system to be 
$\Delta v_{\rm 1D}=$5--7~km\,s$^{-1}$ which is in good agreement with
observations. We assume that energy is dissipated in the disk when
clouds become self-gravitating. Turbulence is generated by 
instabilities involving self-gravitation and is maintained by
the energy input, which is provided by differential rotation via
cloud--cloud interactions. Thus, the dissipative scale
length $l_{\rm diss}$ is equivalent to the size of the largest
self-gravitating clouds. Since we assume clouds to form from local
instabilities due to self-gravitation, which do not depend on the
distance to the galaxy center, $Q$=const. Furthermore, we assume
that the turbulent Reynolds number $Re$ is also independent of the
galactic radius.

\section{The equations}

\subsection{The viscosity prescription \label{sec:vispres}}

In the inertial range of a turbulent medium kinetic energy is
transferred from large scale structures to small scale structures
practically without losing energy. So there is a constant energy flux
from large scales to small scales where the energy is finally dissipated. 
Since the velocity dispersion ($\Delta v \sim 10$ km\,s$^{-1}$) 
within the disk is more important than the
shear, there is no preferred transfer direction. The turbulence is
therefore assumed to be isotropic. In this case the similarity theory
of Kolmogorov applies (see e.g. Landau \& Lifschitz 1959).
The assumption of a universal Kolmogorov equilibrium implies
that the kinetic energy spectrum of the turbulence depends only on 
the energy dissipation rate per unit mass $\epsilon$ and the characteristic
size of the turbulent eddy $l \simeq \frac{1}{k}$, where $k$ is the 
wave number.
The kinetic energy $E(k)$ is related to the mean kinetic energy in
the following way:
\begin{equation}
\frac{1}{2}\langle {\bf u}(x)^{2} \rangle = \int_{0}^{+\infty} E(k) dk\ ,
\end{equation} 
where ${\bf u}$ is the velocity of the medium. Kolmogorov's theory yields 
\begin{equation}
E(k)=C \epsilon^{\frac{2}{3}} k^{-\frac{5}{3}}\ ,
\label{eq:spectrum}
\end{equation} 
where C is a constant of order unity.
Considering a schematic energy spectrum given by
$E(k)=0$, for $k < k_{\rm driv}$ and for $k > k_{\rm diss}$ and by 
Eq.~\ref{eq:spectrum} for $k_{\rm driv} < k < k_{\rm diss}$, 
one derives an expression for the dissipative length scale
$l_{\rm diss} \simeq k_{\rm diss}^{-1}$ and the large scale 
velocity $v_{\rm turb}$:
\begin{equation}
l_{\rm diss}=(\nu^{3}/\epsilon)^{\frac{1}{4}}\ ,
\end{equation}
where $\nu$ is the large scale viscosity due to turbulence;
\begin{equation}
v_{\rm turb}^{2}=\langle {\bf u}^{2} \rangle \simeq \epsilon^{\frac{2}{3}} 
k_{\rm driv}^{-\frac{2}{3}}\ .
\label{eq:v2}
\end{equation}
This leads to a relation between the two length scales and the turbulent
Reynolds number, which is defined by 
$Re \equiv v_{\rm turb}\cdot l_{\rm driv}/\nu$:
\begin{equation}
l_{\rm driv} \simeq Re^{\frac{3}{4}} l_{\rm diss}\ .
\label{eq:redef}
\end{equation}
This turbulent Reynolds number is not equivalent to the
macroscopic Reynolds number defined by e.g. Frank et al. (1992)
\begin{equation}
Re_{\rm macro}=\frac{R\,v_{\rm rot}}{\nu}\ .
\end{equation}
We will show in Sect.~\ref{sec:galaxy} that $Re_{\rm macro} \gg Re$.
In this work we will use the following  viscosity prescription:
\begin{equation}
\nu\,=\,\frac{1}{Re}\,v_{\rm turb}\,l_{\rm driv}\ .
\label{eq:visc}
\end{equation}
In our model the Reynolds number $Re$ defined by Eq.~\ref{eq:redef}
is a free parameter with the restriction $Re \geq 1$. 
This viscosity prescription will be compared to others
in Sect.~\ref{sec:discuss}.

It is formally equivalent to 
the expression for a clumpy accretion disk (Ozernoy et al. 1998)
where the viscosity is
due to cloud--cloud interactions (Eq.~\ref{eq:goldreich}).

We interpret this equivalence
as two different pictures for a turbulent self-gravitating medium.
We prefer the point of view where the whole ISM (all phases) is
taken as one turbulent gas which change phases (i) on turbulent
time scales $t_{\rm turb}=l_{\rm turb}/v_{\rm turb}$ and (ii)
due to external processes (energy output by stars, i.e.
supernovae, UV radiation by O/B stars). Only near the midplane
of the disk can the gas become molecular and self-gravitating which 
leads to a maximum GMC lifetime of 
$\Delta t \sim H_{\rm mol}/v_{\rm turb} \sim 10^{7}$~yr,
where ($H_{\rm mol} \sim 100$~pc Sanders et al. 1985). 
On the other hand the lifetime of a molecular cloud
is approximately given by the crossing time $\Delta t=d_{\rm cl}/v_{\rm turb}$,
where $d_{\rm cl}$ is the cloud size. With a cloud size of 30~pc this results
in $\Delta t \sim 5\,10^{6}$~yr. These values are comparable to
the lifetimes of star-forming molecular clouds given by Blitz \& Shu (1980).
Since these lifetimes are about 10 times smaller than the expected
mean collision time ($\Delta t_{\rm coll} \sim 2\,10^{8}$~yr Elmegreen (1987)),
direct cloud--cloud collisions are very rare and are not important
for the effective viscosity. 

Even with the limited lifetime of molecular clouds, gravitational interactions
always take place between the continuously appearing and disappearing
clouds. These interactions give rise to a collision term in the Boltzmann
equation with an average collision time for gravitational encounters
\begin{equation}
t_{\rm G}=\langle \frac{3v_{\rm turb}^{3}}{4\sqrt{\pi}G^{2}
m_{\rm cl}^{2}n_{\rm cl}}\rangle \ ,
\end{equation}
(Braginskii 1965) where $m_{\rm cl}$ is the cloud mass 
and $n_{\rm cl}$ is the spatial
number density of the clouds. For molecular clouds the turbulent
velocity is approximately independent of the cloud mass 
($v_{\rm turb} \propto m_{\rm cl}^{0.2}$ Larson (1981)), thus
\begin{equation}
t_{\rm G}=\frac{3v_{\rm turb}^{3}}{4\sqrt{\pi}G^{2}\langle
m_{\rm cl}^{2}n_{\rm cl}\rangle}\ ,\ {\rm with}
\end{equation}
\begin{displaymath}
\langle m_{\rm cl}^{2}n_{\rm cl}\rangle = \int m_{\rm cl}^{2} n(m_{\rm cl}) 
\ {\rm d}m_{\rm cl}=
\end{displaymath}
\begin{equation}
6.6\,10^{-6}\ {\rm M_{\odot}^{0.5}\,pc^{-3}}\int m_{\rm cl}^{0.5}\ {\rm d}m_{\rm cl}\ ,
\end{equation}
where we have used the differential number density of molecular clouds
given by Elmegreen (1987).
Using integration boundaries $m_{1}=10^{4}$~M$_{\odot}$,
$m_{2}=10^{6}$~M$_{\odot}$ and $v_{\rm turb}$=10~km\,s$^{-1}$
one obtains $t_{\rm G}\sim 4\,10^{9}$~yr. Assuming a galactic radius
$R$=10~kpc and a rotation velocity of $v_{\rm rot}$=200~km\,s$^{-1}$
gives $\tau = Re \sim 80$.  

Since the dissipative scale length is of the order of the largest
self-gravitating clouds $l_{\rm diss} \sim d_{\rm cl} \sim 30$~pc,
we can calculate the value of the driving wavelength 
$l_{\rm driv}=Re^{\frac{3}{4}} l_{\rm diss} \simeq 800$~pc. 
This is of the same order as the thickness of the distribution of
the atomic gas (Kulkarni \& Heiles 1987, Dickey 1993).
We will show in Sect.\ref{sec:results} that in our model $l_{\rm driv} \sim H$.
The scale height of the molecular disk $H_{\rm mol}$ is smaller,
because of the limited lifetime of molecular clouds.

\subsection{Non Kolomogorov energy spectra}
 
Wada \& Norman (2001) found an energy spectrum of the form $E(k) \propto k^{-2}$
in their 2D hydrodynamical simulations of a supersonic, compressible turbulence
including stellar energy feedback. This power law corresponds to that of a
shock dominated medium. In contrast to a compression
dominated medium, the exponent of the power law for an incompressible medium
is the same in two and three dimensions (Passot et al. 1988).
In this section we assume that  $E(k) \propto k^{-2}$ in three dimensions
In this case, an alternative explanation for this exponent comes from the 
theory of intermittency (Frisch 1995). The open parameter of this model
is the dimension $D$. In the case of intermittent turbulence the energy flux
from scales $\sim l$ to smaller scales is 
\begin{equation}
\epsilon=\frac{v_{l}^{3}}{l} \big(\frac{l}{l_{\rm driv}}\big)^{3-D}=const \ .
\end{equation}
The local spatial average of the energy dissipation can then be expressed as
\begin{equation}
\epsilon_{l}=\frac{v_{l}^{3}}{l}=\epsilon \big(\frac{l}{l_{\rm driv}}\big)^{D-3}\ .
\end{equation}
Within this framework $E(k) \propto k^{-2}$ implies $D=2$ and consequently
$\epsilon_{l}=\epsilon \big(\frac{l}{l_{\rm driv}}\big)^{-1}$, i.e.
the local spatial average of the energy dissipation decreases with decreasing
length scale $l$. The ratio between the driving and the dissipation
length scale is then 
\begin{equation}
\frac{l_{\rm driv}}{l_{\rm diss}}=Re\ ,
\end{equation}
instead of $Re^{\frac{3}{4}}$ in the case of Kolmogorov turbulence.
For a Reynolds number $Re \sim 50$ the difference is about a factor 2.5.
Since the power law exponent of the ISM energy spectrum is not yet
well established, both possibilities must be considered.

\subsection{Angular momentum equation}

In a steady state accretion disk, the mass accretion rate is
\begin{equation}
\dot{M}=2\pi R\,\Sigma\,(-v_{\rm rad})\ ,
\end{equation}
where $v_{\rm rad}$ is the radial velocity. 
The angular momentum equation can be integrated giving
\begin{equation}
\nu \Sigma = -\frac{\dot{M}}{2\pi R} \Omega \big(\frac{\partial \Omega}
{\partial R}\big)^{-1}\ .
\label{eq:ame}
\end{equation} 
Furthermore, we use 
\begin{equation}
\Sigma = \rho\,H  
\label{eq:srh}
\end{equation} 
for the surface density of the disk, where $\rho$ is the mass
density at the midplane.

\subsection{Energy flux conservation \label{sec:efc}} 

The ISM turbulence in the disk is initiated by an instability
involving self-gravitation. The transport of angular momentum
is due to cloud--cloud interactions giving rise to radial mass 
accretion. We assume that the necessary energy input to maintain
the turbulence is the gravitational energy, which is gained when
the ISM is accreted to smaller Galactic radii, i.e. the energy input
is supplied by the Galactic differential rotation. Kinetic energy of the
Galactic rotation is transfered to the turbulent cascade at the driving
wavelength $l_{\rm driv}$ and reaches the dissipative length scale
$l_{\rm diss}$ without loosing energy. The energy per unit time which
is transferred by turbulence is
\begin{equation}
\dot{E}=\rho \nu \int \nabla (v_{\rm turb}^{2})\,{\rm d}{\bf f}
\simeq \rho \nu \frac{v_{\rm turb}^{2}}{l_{\rm driv}} A\ ,
\end{equation}
where the integration is over the area $\int {\rm d}{\bf f}=A$. 
Thus, the energy flux per unit time and unit area is
\begin{equation}
\frac{\Delta E}{\Delta t\,\Delta A}=\rho \nu 
\frac{v_{\rm turb}^{2}}{l_{\rm driv}}\ .
\end{equation}
From standard fluid dynamics, the viscosity $\nu$ generates dissipation
in the disk at a rate
\begin{equation}
\frac{\Delta E}{\Delta t\,\Delta A}= \nu \Sigma (R 
\frac{\partial \Omega}{\partial R})^{2}=-\frac{1}{2\pi}\dot{M}
v_{\rm rot} \frac{\partial \Omega}{\partial R}
\end{equation}
(see e.g. Pringle 1981), where $v_{\rm rot}=\Omega\,R$.
This is valid for $R \gg R_{*}$, where 
$(\partial \Omega/\partial R)_{R=R_{*}}=0$.
Conservation of energy flux leads thus to
\begin{equation}
\rho \nu \frac{v_{\rm turb}^{2}}{l_{\rm driv}}=-\frac{1}{2\pi}\dot{M}
v_{\rm rot} \frac{\partial \Omega}{\partial R}
\label{eq:efc}
\end{equation} 
and relates the mass accretion rate to the turbulence in the disk.

\subsection{The vertical pressure equilibrium}

Since we assume that the sound velocity of the ambient medium 
is smaller than the turbulent velocity dispersion of the clouds,
the only pressure which counterbalances gravitation in the
vertical direction is the turbulent pressure 
$p_{\rm turb}=\rho v_{\rm turb}^{2}$. We distinguish three cases
for the gravitational force in the vertical $z$ direction:
\begin{enumerate}
\item
$M_{\rm d}(R) \leq 0.5\,(H/R)\,M$, where $M_{\rm d}$ is the disk mass and
$M$ is the central mass (dominating central mass);
\item
$\rho \ll \rho_{*}$ and $M_{\rm d}(R) \ll M_{*}(R)$, where $\rho_{*}/M_{*}(R)$ 
is the stellar central density/disk mass 
within a radius $R$ (dominating stellar disk mass);
\item
$\rho \geq \rho_{*}$ and $0.5\,(H/R)\,M(R) < M_{\rm d}(R) < M(R)$, 
where $M(R)$ is the total mass enclosed within a radius $R$ 
(self-gravitating gas disk in $z$ direction). 
\end{enumerate}
For these cases the gravitational pressure $p_{\rm grav}$ has the
following forms:
\begin{enumerate}
\item
$p_{\rm grav}=\rho \Omega^{2} H^{2}$ (see e.g. Pringle 1981);
\item 
$p_{\rm grav}=\pi G \Sigma_{*} \Sigma$, where $\Sigma_{*}$ is the stellar
surface density of the disk (see e.g. Binney \& Tremaine 1987);
\item
$p_{\rm grav}=\pi G \Sigma^{2}$ (Paczy\'{n}ski 1978).
\end{enumerate}
The equilibrium in the vertical direction is given by 
\begin{equation}
p_{\rm grav}=p_{\rm turb}\ .
\label{eq:pp}
\end{equation}

\subsection{Global gravitational stability in $z$ direction}

The basic principles underlying the gravitational instability
of a thin rotating disk can be found in Toomre (1964).
A gaseous disk is locally stable to axisymmetric perturbations, if
\begin{equation}
Q=\frac{v_{\rm turb}\,\kappa}{\pi G\,\Sigma} > 1\ ,
\end{equation}
where $\kappa=\sqrt{R\frac{{\rm d}\Omega^{2}}{{\rm d}R}+4\Omega^{2}}$.
Since in general $\Omega \leq \kappa \leq 2\Omega$, we will use the
following equation:
\begin{equation}
Q = \frac{v_{\rm turb}\,\Omega}{\pi\,G\,\Sigma}\ .
\label{eq:tq}
\end{equation} 
Multiplying the numerator and the denominator of the right hand side 
by $R^{2}$ gives
\begin{equation}
Q=\frac{v_{\rm turb}}{v_{\rm rot}}\frac{M_{\rm tot}}{M_{\rm gas}}\ ,
\label{eq:tqint}
\end{equation}
where $M_{\rm tot/gas}$ is the total enclosed mass and the total enclosed
gas mass at radius $R$. 
Thus for a given velocity dispersion $Q^{-1}$ is proportional to the ratio
of gas mass to total mass.

\section{Results \label{sec:results}}

We have solved the set of equations Eq.~\ref{eq:visc}, Eq.~\ref{eq:ame}, 
Eq.~\ref{eq:srh}, Eq.~\ref{eq:efc}, Eq.~\ref{eq:pp}, and Eq.~\ref{eq:tq}, 
for the three cases:
\begin{enumerate}
\item
dominating central mass,
\item
dominating stellar disk mass,
\item
self-gravitating gas disk in $z$ direction.
\end{enumerate}
We will use $\Omega'=\partial \Omega / \partial R$ with $\Omega' < 0$
in the region of interest.

\subsection{Dominating central mass}

Pressure equilibrium (Eq.~\ref{eq:pp}) leads to
\begin{equation}
H=v_{\rm turb}\Omega^{-1}\ ,
\label{eq:dmc_presseq}
\end{equation}
which is equivalent to the hydrostatic equilibrium where the sound speed
is replaced by the turbulent velocity dispersion.

From energy flux conservation (Eq.~\ref{eq:efc}) and the angular momentum 
equation (Eq.~\ref{eq:ame}) it follows
\begin{equation}
H=\big(R\frac{\Omega'}{\Omega}\big)^{2}l_{\rm driv}\ .
\end{equation}
Thus, the driving wavelength is proportional to the disk height
$l_{\rm driv} \propto H$ and the viscosity prescription reads
$\nu \sim Re^{-1} v_{\rm turb} H$, which is formally equivalent
to the viscosity prescription for clumpy disks of Goldreich \& Tremaine
(1978), Stewart \& Kaula (1980) (see Eq.~\ref{eq:gtsk}), and the
$\alpha$-viscosity (Shakura \& Sunyaev 1973)
with $v_{\rm turb}=c_{\rm s}$ and $Re^{-1}=\alpha$.

Inserting Eq.~\ref{eq:dmc_presseq} in Eq.~\ref{eq:tq} and using
$\rho=\Sigma/H$ gives
\begin{equation}
\rho=\frac{1}{\pi Q}\frac{\Omega^{2}}{G}\ .
\end{equation} 
If we define the critical cloud density for stability against tidal shear
$\rho_{\rm crit}=\pi^{-1}\Omega^{2}G^{-1}$, the Toomre parameter
$Q$ represents the ratio between the critical density and the
volume averaged density in the disk $Q=\rho_{\rm crit}/\rho$. 

\begin{figure}
	\resizebox{\hsize}{!}{\includegraphics{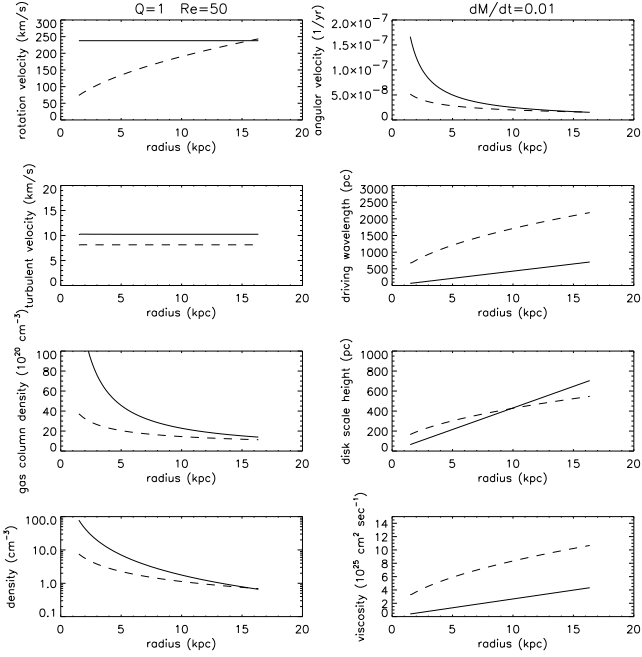}}
      	\caption{ \label{fig:graphs_cdm} 
        Disk parameters for the case of a dominating central mass.
	Solid lines: constant rotation curve. Dashed line: rising
	rotation curve $v_{\rm rot} \propto \sqrt{R}$.
	If $Q$=1 the behaviour of these parameters is the same
	in the case of a selfgravitating gas disk in $z$ direction.
	}
\end{figure}

Inserting the viscosity prescription (Eq.~\ref{eq:visc}) into the angular momentum 
equation (Eq.~\ref{eq:ame}) leads to
\begin{equation}
H=(\frac{1}{2}\,G\,Re\,Q\,\dot{M}\,R\,(-\Omega')\,\Omega^{-4})^{\frac{1}{3}}\ .
\end{equation}
With $\Sigma=\rho / H$ we obtain
\begin{equation}
\Sigma=\rho H = \frac{1}{2^{\frac{1}{3}}\pi} G^{-\frac{2}{3}} Re^{\frac{1}{3}}
Q^{-\frac{2}{3}} \dot{M}^{\frac{1}{3}} R^{\frac{1}{3}} (-\Omega')^{\frac{1}{3}}
\Omega^{\frac{2}{3}}\ .
\label{eq:sigmacmd}
\end{equation}
Inserting Eq.~\ref{eq:sigmacmd} into Eq.~\ref{eq:ame} gives
\begin{equation}
\nu = \frac{1}{2^{\frac{2}{3}}} G^{\frac{2}{3}} Re^{-\frac{1}{3}}
Q^{\frac{2}{3}} \dot{M}^{\frac{2}{3}} R^{-\frac{4}{3}} (-\Omega')^{-\frac{4}{3}}
\Omega^{\frac{1}{3}}\ .
\end{equation}
The turbulent velocity can be calculated using the pressure equilibrium 
(Eq.~\ref{eq:pp}):
\begin{equation}
v_{\rm turb}= H\,\Omega = \big(\frac{1}{2}\,G\,Re\,Q\,\dot{M}\,R\,(-\Omega')\, 
\Omega^{-1}\big)^{\frac{1}{3}}\ .
\end{equation}
If $\Omega' \simeq -\Omega/R$, the viscosity is inversely proportional to the
angular velocity $\nu \propto \Omega^{-1}$ and the turbulent velocity is 
constant. In the case of a Keplerian velocity field due to a point mass
$\nu \propto R^{3/2}$.
The behaviour of the disk parameters ($v_{turb}$, $l_{\rm driv}$, $\Sigma$, $H$, 
$\rho$, and $\nu$) are shown for a constant rotation curve (solid line) and 
a rising rotation curve $v_{\rm rot} \propto \sqrt{R}$ (dashed line) in 
Fig.~\ref{fig:graphs_cdm}. The free parameters $Q=1$, $Re$=50, and
$\dot{M}$=10$^{-2}$~M$_{\odot}$\,yr$^{-1}$ were chosen such that the
disk parameters fit the observed values for the Galaxy (see Sect.~\ref{sec:galaxy}).

\subsection{Dominating stellar disk}

Pressure equilibrium (Eq.~\ref{eq:pp}) together with 
$v_{\rm rot}^{2}=\pi\,G\,\Sigma_{*}\,R$ leads to
\begin{equation}
\frac{H}{R}=\big(\frac{v_{\rm turb}}{v_{\rm rot}}\big)^{2}\ .
\label{eq:h/r}
\end{equation}

Inserting Eq.~\ref{eq:h/r} in the equation for the Toomre parameter
(Eq.\ref{eq:tq}) and using the viscosity prescription (Eq.~\ref{eq:visc}), 
the angular momentum equation (Eq.~\ref{eq:ame}) and energy flux conservation 
(Eq.~\ref{eq:efc}) gives
\begin{equation}
v_{\rm turb}=\big(\frac{1}{2}\,G\,Q\,Re\,\dot{M}\,(-\Omega')\,\Omega^{-2}\big)
^{\frac{1}{2}}\ .
\label{eq:vturbdsd}
\end{equation}
In the case of $\Omega' \simeq \Omega / R$, the turbulent velocity is
inversely proportional to the square root of the rotation velocity
$v_{\rm turb} \propto v_{\rm rot}^{-1/2}$.

Inserting Eq.~\ref{eq:vturbdsd} into Eq.~\ref{eq:h/r} gives the disk scale height
\begin{equation}
H=\frac{1}{2}\,G\,Re\,Q\,\dot{M}\,R^{-1}\,(-\Omega')\,\Omega^{-4}\ .
\label{eq:hdsd}
\end{equation} 
Eq.~\ref{eq:visc}, Eq.~\ref{eq:ame}, and Eq.~\ref{eq:efc} together with Eq.~\ref{eq:hdsd}
lead to the following expression for the driving wavelength
\begin{equation}
l_{\rm driv}=(-\Omega')^{-2}\,\Omega^{2}\,R^{-1}\ .
\end{equation}
From the same equations the density, column density, and viscosity read
\begin{equation}
\rho=\frac{\sqrt{2}}{\pi}G^{-\frac{3}{2}}Q^{-\frac{3}{2}}Re^{-\frac{1}{2}}
\dot{M}^{-\frac{1}{2}}R(-\Omega')^{-\frac{1}{2}}\Omega^{4}\ ,
\end{equation}
\begin{equation}
\Sigma= \frac{1}{\sqrt{2}\pi}G^{-\frac{1}{2}}Q^{-\frac{1}{2}}Re^{\frac{1}{2}}
\dot{M}^{\frac{1}{2}}(-\Omega')^{\frac{1}{2}}\ ,
\label{eq:sigmadsd}
\end{equation}
and
\begin{equation}
\nu=\sqrt{2}G^{\frac{1}{2}}Q^{\frac{1}{2}}Re^{-\frac{1}{2}}
\dot{M}^{\frac{1}{2}}R^{-1}(-\Omega')^{-\frac{3}{2}}\Omega\ .
\label{eq:nudsk}
\end{equation}

\begin{figure}
	\resizebox{\hsize}{!}{\includegraphics{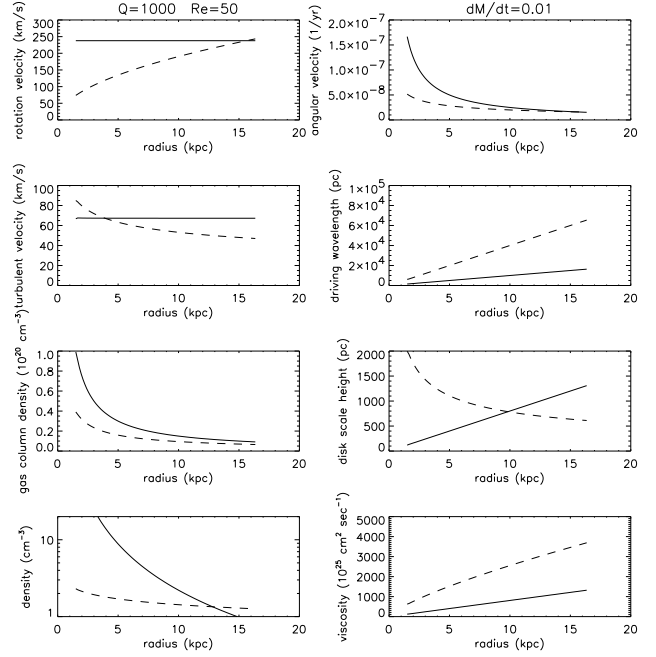}}
      	\caption{ \label{fig:graphs_dsd} 
        Disk parameters for the case of a dominating stellar disk.
	Solid lines: constant rotation curve. Dashed line: rising
	rotation curve $v_{\rm rot} \propto \sqrt{R}$.
	}
\end{figure}

If $\Omega' \simeq -\Omega / R$, the driving wavelength is proportional
to the galactic radius $l_{\rm driv} \propto R$.
Therefore, in the case of a dominating stellar disk, one obtains for the
viscosity prescription (Eq.~\ref{eq:visc}) $\nu \sim Re^{-1}\,v_{\rm turb}\,R$.
In this case, the viscosity is proportional
to the galactic radius divided by the square of the rotation velocity
$\nu \propto R\,v_{\rm rot}^{-1/2}$. Furthermore, if the rotation
curve is flat, i.e. $v_{\rm rot}$=const., $\nu \propto R$.
This behaviour can be observed in Fig.~\ref{fig:graphs_dsd} where
the disk parameters ($v_{\rm turb}$, $l_{\rm driv}$, $\Sigma$, $H$, $\rho$, and
$\nu$) are shown for a constant rotation curve (solid line) and 
a rising rotation curve $v_{\rm rot} \propto \sqrt{R}$ (dashed line)
for $Q$=1000, $Re$=50, and $\dot{M}$=10$^{-2}$~M$_{\odot}$\,yr$^{-1}$.
$Re$, and $\dot{M}$ are the same as for the Galaxy (see Sect.~\ref{sec:galaxy})
and the value of $Q$ is chosen such that the turbulent velocity dispersion
corresponds to that measured of an S0 galaxy ($\sim$50~km\,s$^{-1}$ 
D'Onofrio et al. 1999).

\subsection{Self-gravitating gas in $z$ direction}
 
The pressure equilibrium (Eq.~\ref{eq:pp}) together
with the equation for the Toomre parameter (Eq.~\ref{eq:tq}) leads to
\begin{equation}
\rho=\frac{1}{\pi\,Q^{2}}\frac{\Omega^{2}}{G}\ .
\label{eq:q2}
\end{equation}
The square of the Toomre parameter thus represents the fraction between 
the critical density with respect to tidal shear and the
volume averaged density in the disk $Q^{2}=\rho_{\rm crit}/\rho$.
We want to recall here that for a dominating point mass
$Q=\rho_{\rm crit}/\rho$. 

Inserting Eq.~\ref{eq:q2} into the pressure equilibrium equation 
(Eq.~\ref{eq:pp}) gives
\begin{equation}
H=Q\,\Omega^{-1}\,v_{\rm turb}\ .
\label{eq:hqomega}
\end{equation}

Then, using the angular momentum equation (Eq.~\ref{eq:ame}), it follows
\begin{equation}
l_{\rm driv}=\frac{1}{2}\,Re\,G\,Q\,\dot{M}\,v_{\rm turb}^{-2}\,R^{-1}\,(-\Omega')^{-1}
\label{eq:lsgz}\ .
\end{equation}
From the energy flux conservation (Eq.~\ref{eq:efc}) one obtains
\begin{equation}
v_{\rm turb}=\big(\frac{1}{2}\,G\,Q^{2}\,Re\,\dot{M}\,R\,(-\Omega')
\,\Omega^{-1}\big)^{\frac{1}{3}}\ .
\label{eq:vturb}
\end{equation}
For $\Omega' \simeq -\Omega/R$, the turbulent velocity is constant.
This is what is observed in the nearby galaxy NGC~3938 (van der Kruit
\& Shostak 1982).
 
Inserting this expression in Eq.~\ref{eq:lsgz} leads to
\begin{equation}
\frac{H}{l_{\rm driv}}=Q^{2}R^{2}(-\Omega')^{2}\Omega^{-2}\ .
\label{eq:h/ldriv}
\end{equation}
For $\Omega' \simeq -\Omega/R$, $H/l_{\rm driv} \simeq Q^{2}$.
Thus, the viscosity prescription yields $\nu=Re^{-1}Q^{-2}v_{\rm turb}H$.
Here, $Re$ is not the macroscopic Reynolds number $Re_{\rm macro}=v_{\rm rot}R/\nu$
(see Sect.~\ref{sec:galaxy}).
In the case of $Re=1$ this is the prescription given by Lin \& Pringle (1987a).

Inserting Eq.~\ref{eq:vturb} into Eq.~\ref{eq:hqomega} leads to the disk height 
\begin{equation}
H=\frac{1}{2^{\frac{1}{3}}}Q^{\frac{5}{3}}Re^{\frac{1}{3}}G^{\frac{1}{3}}
\dot{M}^{\frac{1}{3}}R^{\frac{1}{3}}(-\Omega')^{\frac{1}{3}}\Omega^{-\frac{4}{3}}\ .
\label{eq:hsgz}
\end{equation}
With $\Sigma = \rho / H$ one obtains the following expression for the
column density:
\begin{equation}
\Sigma=\frac{1}{2^{\frac{1}{3}}\pi}Q^{-\frac{1}{3}}Re^{\frac{1}{3}}
G^{-\frac{2}{3}}\dot{M}^{\frac{1}{3}}R^{\frac{1}{3}}(-\Omega')
^{\frac{1}{3}}\Omega^{\frac{2}{3}}\ .
\label{eq:sigma}
\end{equation}
Eq.~\ref{eq:visc}, Eq.~\ref{eq:vturb}, Eq.~\ref{eq:h/ldriv}, and Eq.~\ref{eq:hsgz}
lead to
\begin{equation}
\nu=\frac{1}{2^{\frac{2}{3}}}Q^{\frac{1}{3}}Re^{-\frac{1}{3}}G^{\frac{2}{3}}
\dot{M}^{\frac{2}{3}}R^{-\frac{4}{3}}(-\Omega')^{-\frac{4}{3}}
\Omega^{\frac{1}{3}}\ .
\end{equation}
In the case of $\Omega' \simeq -\Omega/R$, the column density is
proportional to and the viscosity is inversely proportional to
the angular velocity: $\Sigma \propto \Omega$, $\nu \propto \Omega^{-1}$.
If in addition the rotation velocity is constant, then $\nu \propto R$.

Comparing this result with Eq.~\ref{eq:nudsk} shows that 
both models with a gravitational potential $\Phi$ due to an extended mass 
distribution give the same viscosity prescription if the rotation velocity 
$v_{\rm rot}=\sqrt{R\,{\rm d}\Phi/{\rm d}R}$ is constant.

On the other hand, the dominating central mass model and the 
self-gravitating gas disk model have the same analytical solutions for $Q=1$
(see Fig.~\ref{fig:graphs_cdm}).

\section{Discussion of the results \label{sec:discuss}}

\subsection{Comparison with previous models of turbulent, selfgravitating gas disks}

Models of turbulent, self-gravitating gas disks can be divided into two different
approaches: (i) The disk is assumed to be quasi continuous and at the edge of
fragmentation ($Q \sim 1$); (ii) The disk is already clumpy and the viscosity is due
to cloud--cloud interactions ($Q \ge 1$). 

We will first discuss the quasi continuous approach.
Paczy\'nski (1978) investigated a self-gravitating disk with a polytropic index
of $\gamma = \frac{4}{3}$ which corresponds to a radiation pressure
dominated disk. Furthermore, he assumed the disk luminosity to be at the
Eddington limit. With the acceleration due to a central mass $M$ and the disk
gas surface density $\Sigma$: $g=GMR^{-3} z+2 \pi G \Sigma$, he obtained 
$\Sigma \propto R^{-3}$ and thus $\nu \propto R^{3}$ for a constant rotation curve.
Lin \& Pringle (1987a) proposed a viscosity prescription based on the Toomre
instability criterion (Toomre 1964): $\nu = Q^{-2}H^{2}R$. They showed that under certain 
conditions this prescription allows a similarity solution $\Sigma \propto R^{-\frac{3}{2}}$.
Their viscosity prescription can be generalized
$\nu \propto \Sigma^{\alpha}R^{-\beta}$ with $\alpha,\ \beta \ge 0$ or
$\nu \propto \Sigma^{\alpha} \Omega^{-\beta}$ for self-gravitating disk
in $z$ and $R$ direction (Saio \& Yoshii 1990).
Shlosman \& Begelman (1989) also considered a disk at the edge of
selfgravitation, i.e. $Q \sim 1$. They obtained $\Sigma \propto v_{\rm turb}v_{\rm rot}R^{-1}
\propto R^{-1}$ for constant turbulent and rotation velocities.
On the other hand, Duschl et al. (2000) made a completely different ansatz:
$\nu \propto \beta v_{\rm rot}R$, with $\beta \ll 1$. They found $\Sigma \propto R^{-1}$
and $\nu \propto R$ for a disk with $v_{\rm rot}=const$.

On the other hand, using a clumpy disk model Silk \& Norman (1981) derived a viscosity 
prescription in assuming that the cooling time for cloud--cloud collisions 
$t_{\rm cool}=l_{0}(v_{\rm turb}\eta)^{-1}$ equals the
viscous time scale $t_{\nu}=R^{2}\nu^{-1}$ for all radii $R$, 
where $\eta$ is the fraction of cloud kinetic 
energy radiated in a collision and $l_{0}$ is the cloud mean free path.
This leads to $\nu=\nu_{0}(R/R_{0}$), where $\nu_{0}=\sqrt{\eta}R_{0}v_{\rm turb}$
and $R_{0}$ is a characteristic length scale. Lynden Bell \& Pringle (1974)
showed that for this viscosity prescription $\Sigma \propto (Rt)^{-1} \exp({-\xi R/t}$),
where $\xi$ is a constant and $t$ is time. Later Shlosman \& Begelman (1987)
also used $t_{\rm coll} \sim t_{\rm visc}$. Ozernoy (1998) and Kumar (1999)
suggested a viscosity prescription based on the collisional Boltzmann equation
(see Sect.~\ref{sec:vispres}) $\nu = \tau v_{\rm turb}^{2} \Omega$, where 
$\tau=(t_{\rm coll}\Omega)^{-1}$.

In this section we will only discuss the self-gravitating gas disk in $z$ direction, 
because only for this set of equation $Q \sim 1$, i.e. the gravitational instability
is active. This point will be further discussed in Sect.~\ref{sec:tgone}.

Whereas the above cited viscosity prescriptions (except Duschl et al. 2000)
use $l_{\rm driv}=H$, the driving wavelength $l_{\rm driv}$ is a priori a
free parameter in our set of equations. It is then defined by the energy
flux conservation (Eq.~\ref{eq:efc}). In terms of the standard 
accretion disk equations (see e.g. Pringle 1981), the ``thermostat'' equation,
i.e. the viscous heating is radiated by the disk , is replaced by the
turbulent energy flux equation (Eq.~\ref{eq:efc}), where the energy gained
through accretion and differential rotation is transported by turbulence to
smaller scales where it is dissipated. The addition of this
energy flux conservation to the ``traditional'' set of disk equations leads
to $l_{\rm driv}=H$. Once this relation is established, the viscosity prescription has the
form of that derived from the collisional Boltzman equation 
with $Re=t_{\rm coll}\Omega$ (see Sect.~\ref{sec:vispres}). On the other hand, our viscosity 
is a factor $Re^{-1}$ smaller than that of Lin \& Pringle (1987a) for $Q \sim 1$.

Thus, a big difference to previous models is that we do not use a disk ``thermostat'', 
i.e. that the viscous heating is radiated by the disk. Radiation is not included
in our model. Duschl et al. (2000)
noticed that the $\alpha$ viscosity together with a pressure term due
to self-gravitation leads to a constant sound velocity. They declared this
unphysical within the framework of a disk ``thermostat''. On the other hand,
if one replaces the sound velocity by the turbulent velocity this is
exactly what is observed in spiral galaxies. 
Since $v_{\rm turb}=const$ we have $H/R=const.$, and consequently $\nu \propto R$ 
and $\Sigma \propto R^{-1}$.

\subsection{Why can the Toomre parameter be greater than 1? \label{sec:tgone}}

Whereas the two approaches of a continuous and a clumpy disk can be unified,
this can not be done for a disk with $Q > 1$. In this case the disk is
not globally gravitational unstable (Toomre 1964). The turbulent velocity is so high,
or the surface density is so low that the disk is globally gravitationally
stable. However, we can imagine two possibilities for the formation of $Q > 1$ disks:

(i) In Sect.~\ref{sec:galaxy} we will show that the mass accretion rate
within the disk is much smaller than the star formation rate. Thus, starting
from a $Q \sim 1$ disk with star formation, $Q$ will increase with time
(if there is no external mass infall). This could be the case for S0 galaxies.

(ii) The gas which falls into the given gravitational potential is already
clumpy. This could be the case for the Circumnuclear Disk in the Galactic Center 
(Vollmer \& Duschl 2000).

For these scenarios the above models for a dominating central mass and
a dominating stellar disk are valid. We are the first to solve these sets of equations.

We will now show that $Q=const > 1$ can be understood in terms of star formation.
With the critical density for tidal disruption $\rho_{\rm crit}=\Omega^{2}/(\pi\,G)$
and using $\dot{\Sigma}_{*}=Re^{-1} \Sigma \Omega$ (Eq.~\ref{eq:silk}) one obtains:
\begin{equation}
\dot{\rho}_{*}=\frac{1}{Re\,Q}\frac{\rho_{\rm crit}}{t_{\rm H}}\ ,
\end{equation}
where $t_{\rm H}=H/v_{\rm turb}$ is the vertical crossing time.
Thus, $Q=const > 1$ together with $Re=const > 1$ means that the
star formation rate is proportional to the critical density with respect
to tidal shear divided by the vertical crossing time.

The detailed comparison between observations and our model will show
if $Q > 1$ models are useful to describe clumpy disks with a low gas mass or high velocity 
dispersion.

\subsection{Energy dissipation due to selfgravitation}

If turbulence is dissipated by self-gravitation, the energy dissipation rate due
to self-gravitation of the disk in $z$ direction $\epsilon_{\rm sg}$ must equal the constant, 
turbulent energy dissipation rate per unit mass $\epsilon_{\rm turb}$.

The energy due to self-gravitation of a gaseous disk is 
\begin{equation}
\Delta E = p\,\Delta A\,H = \pi G \Sigma^{2}\,\Delta A\,H = \pi G \Sigma M_{\rm gas} H\ ,
\end{equation} 
where $\Delta A$ is the disk surface. The growth rate of the gravitational
instability in $z$ direction is approximately the free fall time
$\Delta t \sim \sqrt{H/(G\Sigma)} = \sqrt{(G\rho)^{-1}}$.
Thus, one obtains for the energy dissipation rate per unit mass due to self-gravitation
\begin{equation}
\epsilon_{\rm sg} = \frac{\Delta E}{\Delta M\,\Delta t}=\pi G \Sigma H \sqrt{G \rho}\ .
\end{equation}
Using Eq.~\ref{eq:q2} and Eq.~\ref{eq:hqomega} it can be shown that
\begin{equation}
\epsilon_{\rm sg} \sim \epsilon_{\rm turb} = \frac{v_{l_{\rm diss}}^{3}}{l_{\rm diss}}=
\frac{v_{\rm turb}^{3}}{l_{\rm driv}}=\frac{v_{\rm turb}^{3}}{H}\ .
\end{equation}
In order to estimate the mass of a self-gravitating cloud, we use
$\Delta E \sim N_{\rm cl} \frac{3}{5}M_{\rm cl}^{2}G/l_{\rm diss}$, where $N_{\rm cl}$
is the number of selfgravitating clouds and $M_{\rm cl}$ is the cloud mass.
With the free fall time of the clouds $\Delta t \sim \sqrt{(G\rho)^{-1} \Phi_{\rm V}}$
this leads to
\begin{equation}
\frac{\Delta E}{\Delta M\,\Delta t}=\frac{3}{5}Re^{-\frac{3}{4}}H^{-1}\sqrt{G\rho \Phi_{\rm V}^{-1}}=\epsilon_{\rm turb}\ ,
\end{equation}
where $\Phi_{\rm V}$ is the volume filling factor (see Sect.~\ref{sec:vff}) and
$\Delta M = N_{\rm cl} M_{\rm cl}$. 
Using the expression for $\Phi_{\rm V}$ derived in Sect.~\ref{sec:vff} and 
Eq.~\ref{eq:q2} gives
\begin{equation}
M_{\rm cl} \sim \frac{5\sqrt{\pi}}{3}G^{-1}Re^{-\frac{5}{4}}Q^{-1}
\Omega^{-1}v_{\rm turb}^{3}\ .
\end{equation}
With $Q=1$, $Re$=50, $\Omega=2\,10^{-8}$~yr$^{-1}$, and $v_{\rm turb}$=10~km\,s$^{-1}$
one obtains $M_{\rm cl} \sim 2\,10^{5}$~M$_{\odot}$. This is consistent with
the mass of GMCs.

In the present paper we have assumed that the potential energy gained through
accretion is dissipated locally. This is true only in the absence of radial
energy transport, which can be checked a posteriori.
The energy dissipation rate due to self-gravitation of the disk 
in $z$ direction is
\begin{equation}
\delta_{\rm sg}=\frac{\Delta E}{\Delta A \Delta t} \sim \frac{\pi G \Sigma^{2} H}{t_{\rm H}}\ .
\end{equation} 
The radial energy flux is given by
\begin{equation}
\delta_{\rm r} = \Sigma v_{\rm r} \frac{\partial e}{\partial R}-\Sigma v_{\rm turb}^{2}
R^{-2} \frac{\partial}{\partial R}(R^{2}v_{\rm r})\ ,
\end{equation} 
where $v_{\rm r}=-\dot{M}/(2 \pi R \Sigma)$ is the radial velocity
and $e$ the specific internal energy.
For supersonic turbulence in clumpy disks $e$ is dominated by the specific kinetic
energy $v_{\rm turb}^{2}$. Since $v_{\rm turb}=const$, $\partial e / \partial R \sim 0$.
With $v_{\rm r}=const$ one obtains
\begin{equation}
\delta_{\rm r} = -2 \Sigma v_{\rm turb}^{2} v_{\rm r} R^{-1}=-2 \rho v_{\rm turb}^{2} v_{\rm r}
H R^{-1}\ .
\end{equation}
Using $t_{\rm H}=H/v_{\rm turb}$ and Eqs.~\ref{eq:q2}, \ref{eq:vturb}, \ref{eq:hsgz}, \ref{eq:sigma} 
one obtains
\begin{equation}
\frac{\delta_{\rm r}}{\delta_{\rm sg}}=\sqrt{\frac{3}{32}}2^{-\frac{2}{3}}\pi
Q^{\frac{4}{3}}Re^{-\frac{1}{3}}G^{\frac{2}{3}}\dot{M}^{\frac{2}{3}}v_{\rm rot}^{-2}\ .
\end{equation}
With $Q$=1, $Re$=50, and $\dot{M}=10^{-2}~$M$_{\odot}$yr$^{-1}$ (see Sect.~\ref{sec:galaxy})
this leads to
\begin{equation}
\frac{\delta_{\rm r}}{\delta_{\rm sg}} \sim \frac{2}{(v_{\rm rot} [{\rm km\,s}^{-1}])^{2}}
\ll 1\ .
\label{eq:deltarsg}
\end{equation}
The radial advective energy flux is much smaller than the energy gain by mass accretion
and can therefore be neglected as long as Eq.~\ref{eq:deltarsg} holds.

We can conclude here that the energy flux transported to 
small scales by turbulence has the same order of magnitude as the energy dissipation 
rate due to selfgravitation of the gaseous disk.
Moreover, the potential energy gained through accretion is dissipated locally at small scales.
Since our model fits observations for a reasonable choice of parameters,
we conclude that, based on energy flux conservation, it is possible that turbulence 
is generated and dissipated through gravitational instabilities, and maintained by 
the energy input due to mass inflow and differential rotation (Sect.~\ref{sec:efc}).

\section{Implications}

\subsection{Volume filling factors \label{sec:vff}}

In this Section we compare the crossing time of a turbulent cloud to the
gravitational free fall time in order to derive an expression for the
volume filling factor as a function of the Reynolds number $Re$ and the
Toomre parameter $Q$. We assume turbulence with a Kolmogorov spectrum
for $l_{\rm diss} \leq l \leq l_{\rm driv}$. This implies:
$l_{\rm diss}=Re^{-\frac{3}{4}}l_{\rm driv}$ and 
$v_{\rm cl}=Re^{-\frac{1}{3}}v_{\rm turb}$, where $v_{\rm cl}$ is the turbulent 
velocity of a cloud of size $l_{\rm diss}$.

The characteristic turbulent time scale of clouds whose size is comparable 
to the dissipative length scale is
\begin{equation}
t_{\rm l}=l_{\rm diss}/v_{\rm cl}\ .
\end{equation}
The gravitational free fall time is given by
\begin{equation}
t_{\rm ff}=\sqrt{\frac{3\pi}{32G\rho_{\rm cl}}}\ ,
\end{equation}
where $\rho_{\rm cl}$ is the density of a single cloud, which is related
to the overall disk density $\rho$ by the volume filling factor $\phi_{\rm V}$:
$\rho_{\rm cl}=\phi_{\rm V}^{-1}\rho$.

The clouds are self gravitating for $t_{\rm l}=t_{\rm ff}$. Inserting the
expressions given in Sect.~\ref{sec:results} leads to:
\begin{enumerate}
\item
dominating central mass:\\
$\phi_{\rm V} = \frac{32}{3\pi^{2}}\,Q^{-1}Re^{-1}R^{-4}(-\Omega')^{-4} \Omega^{4}$.
\item
dominating stellar disk mass:\\
$\phi_{\rm V} = \frac{64 \sqrt{2}}{3\pi^{2}}G^{-\frac{3}{2}}Re^{-\frac{5}{2}}Q^{-\frac{5}{2}}
\dot{M}^{-\frac{3}{2}}R^{-1}(-\Omega')^{-\frac{11}{2}}\Omega^{10}=$\\
$=\frac{32}{3\pi^{2}}Re^{-1}Q^{-1}R^{-4}(-\Omega')^{-4}\Omega^{4}\big(\frac{v_{\rm rot}}{v_{\rm turb}}\big)^{3}$.
\item
self-gravitating gas disk in $z$ direction:\\
$\phi_{\rm V} = \frac{32}{3\pi^{2}}\,Q^{-4}Re^{-1}R^{-4}(-\Omega')^{-4}\Omega^{4}$.
\end{enumerate}

For a non Kolmogorov spectrum $E(k) \propto k^{-2}$, the volume filling factors have to be 
multiplied by $Re^{-1}$:  $\Phi_{\rm V}^{\rm non K.} = Re^{-1} 
\Phi_{\rm V}^{\rm K.}$.

\subsection{Bondi--Hoyle accretion limit}

We will now consider the limit $\phi_{\rm V}=Q=Re\sim 1$, which means
$M_{\rm gas} \sim M_{\rm tot}$ and $v_{\rm turb} \sim v_{\rm rot}$.
This implies $H \sim R$, i.e. this is the limit for a spherical configuration.
It is assumed that $\partial \Omega/\partial R \simeq -\Omega/R$.
The viscosity prescription (Eq.~\ref{eq:visc}) together with
energy flux conservation equation (Eq.~\ref{eq:efc}) gives
\begin{equation}
\frac{1}{Re}\rho v_{\rm turb}^{3}=\frac{\dot{M}}{2\pi}v_{\rm rot}\frac{\Omega}{R}\ .
\end{equation}
Setting $v_{\rm turb}=v_{\rm rot}$, $Re=1$ and using $v_{\rm rot}^{2}=MG/R$ leads to
\begin{equation}
\dot{M}=\frac{2\pi G^{2}M^{2}\rho}{v_{\rm rot}^{3}}\ .
\end{equation}
This is equivalent to the Bondi--Hoyle accretion rate (Bondi \& Hoyle 1944), 
which is given by
\begin{equation}
\dot{M}_{\rm B-H}=\frac{2.5\pi G^{2}M^{2}\rho}{v_{\rm turb}^{3}}
\end{equation}
for $v_{\rm turb}=v_{\rm rot}$.

\subsection{The molecular fraction}

In order to derive an expression for the molecular fraction of the gas in the disk,
we compare the crossing time of the turbulent layer $t_{\rm turb}$ and the 
H--H$_{2}$ transition time scale $t_{\rm H_{2}}=\alpha/\rho$ (Hollenbach \& Tielens 1997).
We define the molecular fraction here as $f_{\rm mol}=t_{\rm turb}/t_{\rm H_{2}}$.

For the three different cases we obtain using a Kolmogorov spectrum:
\begin{enumerate}
\item
dominating central mass:\\
$f_{\rm mol}=1/(\pi\alpha)\,Q^{-1}Re^{-\frac{1}{2}}G^{-1}R^{-2}(-\Omega')^{-2}\Omega^{3}$,
\item
dominating stellar disk mass:\\
$f_{\rm mol}=2/(\pi\alpha)\,G^{-2}Q^{-2}Re^{-\frac{3}{2}}\dot{M}^{-1}(-\Omega')^{-3}\Omega^{7}=$\\
$=2/(\pi\alpha)\,G^{-2}Q^{-2}Re^{-\frac{3}{2}}\dot{M}^{-1}R^{-3}(-\Omega')^{-3}v_{\rm rot}^{3}\Omega^{4}$,
\item
self-gravitating gas disk in $z$ direction:\\
$f_{\rm mol}=1/(\pi\alpha)\,Q^{-3}Re^{-\frac{1}{2}}G^{-1}R^{-2}(-\Omega')^{-2}\Omega^{3}$.
\end{enumerate}

For a non Kolmogorov spectrum $E(k) \propto k^{-2}$, the molecular fractions have to be 
multiplied by $Re^{-\frac{1}{2}}$:  $f_{\rm mol}^{\rm non K.} = Re^{-\frac{1}{2}} 
f_{\rm mol}^{\rm K.}$.

In the following we will only treat the case of a self-gravitating gas disk in
$z$ direction, because it represents the most realistic description of
normal field spiral galaxies. We will motivate this choice in Section~\ref{sec:sftf}.

For a galaxy with a constant rotation velocity the molecular gas surface density 
is given by
\begin{equation}
\Sigma_{\rm H_{2}}=\frac{1}{2^{\frac{1}{3}}\pi^{2}\alpha}Q^{-\frac{10}{3}}Re^{-\frac{1}{6}}
G^{-\frac{5}{3}}\dot{M}^{\frac{1}{3}}\Omega^{2}\ . 
\end{equation}
The total integrated molecular gas mass is then
\begin{equation}
M_{\rm H_{2}}=\pi \Sigma_{\rm H_{2}} R^{2}=\frac{1}{2^{\frac{1}{3}}\pi\alpha}Q^{-\frac{10}{3}}
Re^{-\frac{1}{6}}G^{-\frac{5}{3}}\dot{M}^{\frac{1}{3}}v_{\rm rot}^{2}\ , 
\end{equation}
which does not depend on the galactic radius.

With these results we can now calculate the ratio of molecular to atomic gas in
spiral galaxies 
\begin{equation}
M_{\rm H_{2}}/M_{\rm HI}=M_{\rm H_{2}}/(M_{\rm gas}^{\rm tot}-M_{\rm H_{2}})\ ,
\end{equation}
where $M_{\rm gas}^{\rm tot}=\pi \Sigma R^{2}$ is the total gas mass.

\subsection{Star formation and the Tully--Fisher relation \label{sec:sftf}}

The star formation rate per unit area $\dot{\Sigma}_{*}$ is usually described by a 
Schmidt law (Schmidt 1959) of the form 
\begin{equation}
\dot{\Sigma}_{*} \propto \Sigma_{\rm gas}^{N}\ ,
\end{equation}
where $N \sim 1.4$ (Kennicutt 1998). An alternative description of the star
formation rate was proposed by Elmegreen (1997) and Silk (1997).
They suggested that it scales with the ratio of the gas density to the average
orbital time scale
\begin{equation}
\label{eq:silk}
\dot{\Sigma}_{*} = 0.017 \Sigma_{\rm gas} \Omega\ .
\end{equation}
This parameterization provides a fit that is nearly as good as the Schmidt law.

On the basis of our model, we suggest that the star formation time scale
is $\tau_{\rm SF}=Re\,\Omega^{-1}$, which is the interaction time scale
between the clouds. This corresponds to the time scale
of gravitational encounters between the clouds. We thus find $Re\sim 60$, which
is only a factor 2 higher than our estimate based on the disk height and
the cloud size. In the case of a constant rotation velocity our star formation 
prescription yields in the case of selfgravitation in $z$ direction:
\begin{equation}
\dot{\Sigma}_{*}=\frac{1}{2^{\frac{1}{3}}\pi}Q^{-\frac{1}{3}}Re^{-\frac{2}{3}}
G^{-\frac{2}{3}}\dot{M}^{\frac{1}{3}}\Omega^{2}\ .
\label{eq:sfr}
\end{equation}
Thus both, the star formation rate per unit area and the molecular gas density,
are proportional to the square of the angular velocity $\Sigma_{\rm H_{2}} \propto
\Omega^{2}\ ,\dot{\Sigma}_{*} \propto \Omega^{2}$. Both profiles have thus
the same dependence on the galactic radius. This is what is found when comparing
CO and H$\alpha$, radio continuum, B band luminosity profiles of most of the Virgo 
cluster galaxies observed by Kenney \& Young (1989). More recently, Rownd \& Young (1999) 
combined H$\alpha$ imaging and CO--line observations to derive radial star formation
efficiencies for a sample of cluster and field spiral galaxies. They found that
$\dot{\Sigma}_{*} / \Sigma_{\rm H_{2}} = const$, which is consistent with
our findings. It is worth noting, that in the framework of our model this
proportionality only exists between the star formation rate and the molecular gas mass
and {\it not} the total gas mass.

Whereas the star formation laws of Kennicutt (1998) or Silk (1997) depend explicitly
on the gas surface density, our prescription only depends on the angular velocity $\Omega$.
The proportionality factor depends on $\dot{M}$ and the free parameters $Re$, $Q$.
The results of Rownd \& Young (1999) imply that the combination of these
parameters or all parameters individually are the same for all spiral galaxies. 
The prescription of Silk (1997) is equivalent to ours 
$\dot{\Sigma}_{*} = \xi \Sigma \Omega$.
We identify $\xi=Re$. The observationally derived $\xi$ is consistent
with the definition of $Re=(l_{\rm driv}/l_{\rm diss})^{\frac{3}{4}}$.
For the comparison between the Kennicutt and Silk law we refer to Kennicutt(1998).

The integrated star formation rate is 
\begin{equation}
\dot{M}_{*}=\frac{Q}{2}\big(\frac{v_{\rm rot}}{v_{\rm turb}}\big)^{2}\dot{M}=
\frac{1}{2^{\frac{1}{3}}}Q^{-\frac{1}{3}}Re^{-\frac{2}{3}}
G^{-\frac{2}{3}}\dot{M}^{\frac{1}{3}}v_{\rm rot}^{2}\ .
\label{eq:mdot}
\end{equation}
Thus the integrated star formation rate is proportional to the square
of the rotation velocity as the total amount of molecular gas. 
Observationally, Devereux \& Hameed (1997) found that the ratio of
star formation rate to molecular gas ($L_{\rm FIR}/M({\rm H_{2}})$)
is independent of the morphological type of a spiral galaxy, thus
$\dot{M}_{*} \propto M_{\rm H_{2}}$. 

The blue Tully--Fisher relation is dominated by light associated with current
star formation. In general, the galaxy luminosity is proportional to the
rotation velocity at a certain power $L \propto v_{\rm rot}^{\gamma}$.
Compilations of Tully--Fisher data for available samples find that 
$\gamma$=2.1--2.2 for the B band with a systematic increase with increasing
wavelength (Silk 1997). Our model would ideally predict 
$\dot{M}_{*} \propto v_{\rm rot}^{2}$, i.e. $\gamma=2$. Since the B band
luminosity is provided by a mixture of stars of different ages, we would expect
that $\gamma$ is slightly higher than predicted by our model.

We will now calculate the integrated star formation rate for the case of a galaxy
with dominating stellar disk. Inserting the
expression for the volume filling factor for a constant rotation velocity disk
$\phi_{\rm V} \simeq 
\frac{4}{3\pi}\,Q^{-1}Re^{-1}(v_{\rm rot}/v_{\rm turb})^{3}$ and Eq.~\ref{eq:sigmadsd}
into the expression for the integrated star formation rate 
$\dot{M}_{*}=\pi\Sigma R^{2}Re^{-1}\Omega$ yields
\begin{equation}
\dot{M}_{*}=\frac{v_{\rm rot}}{v_{\rm turb}}\dot{M}=
(\frac{3\pi}{4}Re\,Q\,\Phi_{\rm V} )^{\frac{1}{3}} \dot{M}\ .
\end{equation}

Taking $\phi_{\rm V}=10^{-2}$, $Re=50$, $\dot{M}=10^{-2}$~M$_{\odot}$yr$^{-1}$,
and $v_{\rm rot}=200$~km\,s$^{-1}$ gives $v_{\rm turb} \sim 50$~km\,s$^{-1}$ and
$\dot{M}_{*} \sim 4\,10^{-2}$~M$_{\odot}$yr$^{-1}$. Thus, the integrated
star formation rate is comparable to the mass accretion rate. Such low
integrated star formation rates are only observed for S0 galaxies
(Kennicutt 1998). The Toomre parameter for such a disk is $Q \sim 7\,10^{3}$.
For $\phi_{\rm V}=1$, $Q\,Re^{\frac{5}{6}}=4(v_{\rm rot}/v_{\rm turb})^{3} \geq 10^{3}$.
This is not the case for normal field spirals. Thus, for most of the field spiral 
galaxies, the case of a self-gravitating disk in $z$ direction applies.

\subsection{Low surface brightness galaxies}

Low surface brightness (LSB) galaxies have low density/column density gas
but a high gas fraction $M_{\rm gas}/M_{\rm tot}$ (de Blok 1999).
These LSB galaxies can be either dwarf galaxies with small rotation velocities
(see e.g. Walter \& Brinks 1999) or large galaxies with high rotation velocities
(see e.g. Pickering et al. 1997). Both have small angular velocities
$\Omega=v_{\rm rot}/R$. Since our model yields $\Sigma_{\rm gas} \propto \Omega$ and
$\rho \propto \Omega^{2}$, a small angular velocity implies a low gas column density. 
Thus, the observed small gas density/column density of LSB galaxies can be 
partly due to their small angular velocity.

\section{Application to the Galaxy \label{sec:galaxy}}

\begin{figure}
	\resizebox{\hsize}{!}{\includegraphics{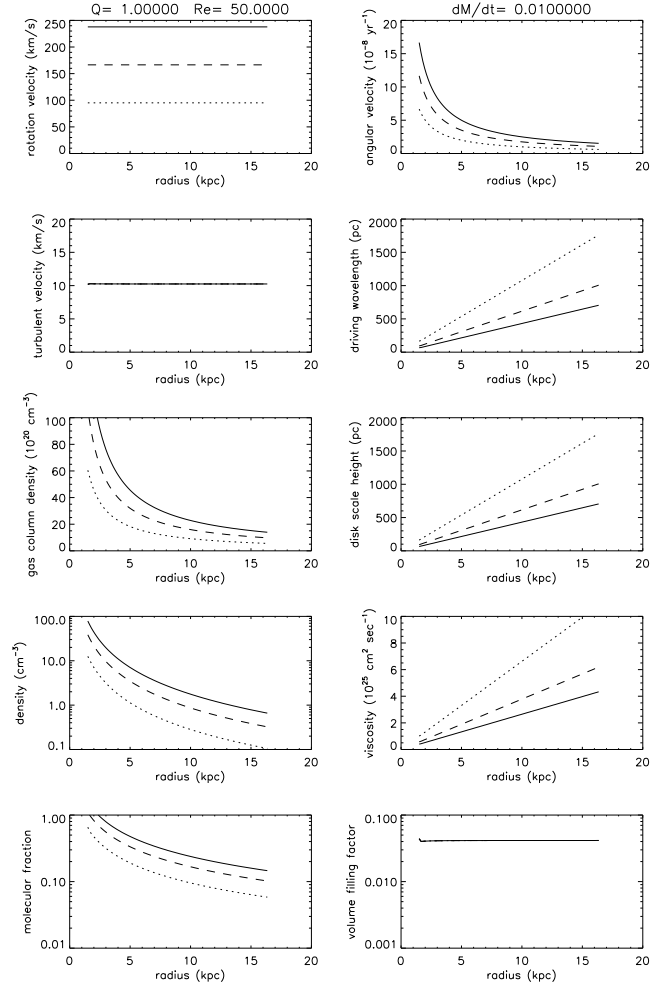}}
      	\caption{ \label{fig:graphs_galaxy} 
        	Radial profiles of the rotation velocity, angular velocity,
		turbulent velocity, driving wavelength, gas column density, 
		disk scale height, gas density, viscosity, molecular fraction, 
		and volume filling factor for three different rotation velocities
		(solid line: $v_{\rm rot}=$250~km/,s$^{-1}$, dashed line:
		$v_{\rm rot}=$175~km/,s$^{-1}$, dotted line: 
		$v_{\rm rot}=$100~km/,s$^{-1}$).
	}
\end{figure}

As motivated in Sect.~\ref{sec:sftf} the model of a vertically self-gravitating 
gas disk describes the gas disks of spiral galaxies.
We will use the local parameter of the ISM in the solar neighbourhood 
given by Binney \& Tremaine (1987): $\rho=0.042$~M$_{\odot}$pc$^{-3}$,
$\Sigma=5.3$~M$_{\odot}$pc$^{-2}$, $\Omega=2.62\,10^{-8}$~yr$^{-1}$ and $R$=10~kpc.
Furthermore, we have shown in Section~\ref{sec:vispres} and 
Section~\ref{sec:sftf} that $Re \sim$ 60--80.
The rotation curve is assumed to be constant. Eq.~\ref{eq:q2} leads to
$Q \sim 1$. On the other hand, the definition of the $Q$ parameter (Eq.~\ref{eq:tq})
would yield $v_{\rm turb} \sim 3$~km\,s$^{-1}$. This equation would thus 
require $Q \sim 3$ in order to match the gas velocity dispersion
of $v_{\rm turb}$=10~km\,s$^{-1}$ found by van der Kruit \& Shostak (1982). 
We will here adopt\\ 
$Q=1$,\\
$Re=50$,\\ 
$\alpha=10^{7}$~yr\,M$_{\odot}$pc$^{-3}$
(this corresponds to a molecular transition time scale of 
$t_{\rm mol}\sim 5\,10^{8}/n$~yr, where $n$ is the particle density),\\
$R$=10~kpc, and\\
$v_{\rm turb}$=10~km\,s$^{-1}$.\\
Eq.~\ref{eq:vturb} then leads to $\dot{M}\sim 10^{-2}$~M$_{\odot}$yr$^{-1}$.
Thus, we suggest that the Galactic mass accretion rate is
$\dot{M}\ \sim\ 10^{-2}$~M$_{\odot}$yr$^{-1}$. 
Fig.~\ref{fig:graphs_galaxy} shows the derived turbulent velocity, driving
wavelength, gas column density, disk scale height, gas density, viscosity,
molecular fraction, and volume filling factor for three different
rotation velocities.
For the model of the Galaxy, we have adopted\\
$v_{\rm rot}$=250~km\,s$^{-1}$.

\begin{figure}
	\resizebox{\hsize}{!}{\includegraphics{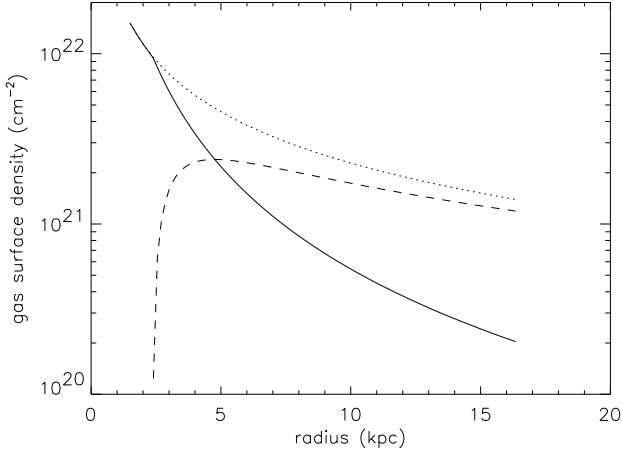}}
      	\caption{ \label{fig:profiles} 
        	Radial profiles of the total gas (dotted line), the molecular (solid line), 
		and the atomic (dashed line) gas surface density.
	}
\end{figure}

This results in a total gas mass of
$M_{\rm gas}^{\rm tot}=5.4\,10^{9}$~M$_{\odot}$, a total molecular gas mass
$M_{\rm H_{2}}=1.7\,10^{9}$~M$_{\odot}$, a total atomic gas mass
$M_{\rm HI}=3.7\,10^{9}$~M$_{\odot}$, and $M_{\rm H_{2}}/M_{\rm HI}=0.3$.
Scoville \& Sanders (1987) give $M_{\rm HI} \sim 2.5\,10^{9}$~M$_{\odot}$, 
$M_{\rm H_{2}} \sim 0.7\,10^{9}$~M$_{\odot}$ (we have used the conversion factor
$X=10^{20}$~cm$^{-2}$(K\,km\,s$^{-1}$)$^{-1}$ of Digel et al. 1996) 
and thus $M_{\rm H_{2}}/M_{\rm HI}\sim 0.3$.
Our model calculations are consistent within a factor of 2 with the 
absolute values of the total gas mass
and the observed ratio of molecular to atomic gas in the Galaxy.
Figure~\ref{fig:profiles} shows the radial profiles of the
total gas (dotted line), the molecular (solid line), and the
atomic (dashed line) gas surface density for $v_{\rm rot}=250$~km\,s$^{-1}$
and $\alpha = 7\,10^{6}$~yr\,M$_{\odot}$pc$^{3}$. 
These profiles well resemble that found by Kenney \& Young (1989) (their Fig.~12).
Honma et al. (1995) examined the H{\sc i} and H$_{2}$ radial distributions
of four face--on spiral galaxies. They found that the gas phase transition occurs 
suddenly within a narrow range of radius. This transition is explained using a
theory whose main parameters are the ISM pressure, the UV radiation field, and
the metallicity. It was found that the metallicity gradient is most crucial 
for the formation of a narrow range transition. Our model (Fig.~\ref{fig:profiles})
also shows a sharp phase transition. For a given $Re$, $Q$, and $\dot{M}$,
the location of this transition region depends mainly on the formation time scale $\alpha$
of H$_{2}$ from H{\sc i}. If $\alpha$ depends on metallicity (and that is
what one would expect), the two models are in good agreement, because radiation,
which is not included in our model, does not play a major r\^{o}le.

The disk height equals the driving wavelength $H \sim l_{\rm driv}\sim 400$~pc,
and the viscosity is $\nu\sim 8\,10^{-5}$~pc$^{2}$yr$^{-1}=2\,10^{25}$cm$^{2}$s$^{-1}$.
Thus, the macroscopic Reynolds number is $Re_{\rm macro}=v_{\rm rot}R/\nu
\sim 3\,10^{4} \gg Re$. This value of the driving wavelength is only
a factor of 2 larger than that found by Wada \& Norman (2001) in their
2D hydrodynamical simulations.

The derived mass accretion rate is much smaller than the star formation rate
(Eq.~\ref{eq:mdot} yields $\dot{M}_{*} \sim 3$~M$_{\odot}$yr$^{-1}$ compared to
$\dot{M}_{*} \sim 3-5$~M$_{\odot}$yr$^{-1}$ as suggested by the observations of
other Sbc spirals; Tammann et al. 1994; Cappellaro et al. 1997).
The current viscous time scale $t_{\nu} \sim R^{2}/\nu \sim 10^{12}$~yr
is larger than a Hubble time. Since the viscous time scale is given by 
\begin{equation}
t_{\nu} \sim \frac{R^{2}}{\nu} \propto \dot{M}^{-\frac{2}{3}}R^{2}\Omega \sim
\dot{M}^{-\frac{2}{3}} v_{\rm rot}R\ ,
\end{equation}
viscous evolution has taken place at a time when the mass accretion
rate was higher and/or the disk was smaller and/or rotation velocity was smaller. 
After this evolution the gas distribution reached an equilibrium state
that is described by our model. Lin \& Pringle (1987b) and later Saio \& Yoshii (1990)
have shown that one obtains an exponential stellar disk if $t_{\nu} \sim t_{*}$.
In our model this is only the case for a dominating stellar disk, where
$\dot{M_{*}} \sim \dot{M}$. However, we have to stress here that $\dot{M}$ is the
mass accretion rate in the disk. If there is also external accretion present,
this formally increases $t_{*}$, which can lead to $t_{\nu} \sim t_{*}$.

\section{Morphological type versus mass}

In this Section we investigate the link between the morphological classification
of galaxies and their mass on the basis of our model for vertically
selfgravitating gas disks. In general, disk galaxies of later types (Sc--Scd)
are smaller and less massive than galaxies of earlier type (Sa--Sab)
(Roberts \& Haynes 1994). Some of the properties which lead to the morphological
classification could thus be due to the galaxies' gravitational potential
as proposed by Gavazzi et al. (1996).

\subsection{The velocity dispersion and the fraction of gas to total mass}

It is a surprising fact that galactic gas disks all have dispersion velocities between 
8 and 10~km\,s$^{-1}$ (see e.g. Freeman 1999). Assuming a constant rotation
velocity Eq.~\ref{eq:vturb} implies that $Q^{2}\,Re\,\dot{M}$ is approximately
constant for spiral galaxies of all types and all luminosities.
Furthermore, if the galaxy forms still actively stars $Q$ should be $\sim$1.
Thus $Re\,\dot{M}$ can only vary by a factor $\sim$10.

From Eq.~\ref{eq:tqint} we find $M_{\rm gas}/M_{\rm tot} = Q^{-1} v_{\rm turb}/v_{\rm rot}$.
Since the gas velocity dispersion is almost constant, 
$M_{\rm gas}/M_{\rm tot} \propto v_{\rm rot}^{-1}$. Galaxies with high rotation velocities
have thus smaller gas to total mass ratios. This trend is also
observed in the Hubble sequence: early type spirals have larger rotation
velocities and smaller gas to total mass ratios (Roberts \& Haynes 1994).

\subsection{The fraction of molecular to atomic gas}

\begin{figure}
	\resizebox{\hsize}{!}{\includegraphics{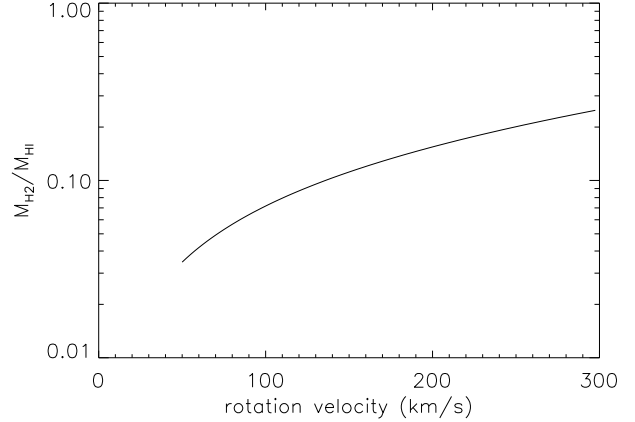}}
      	\caption{ \label{fig:young_knezek} 
        Ratio of molecular to atomic gas mass as a function of the galaxy rotation velocity.
	}
\end{figure}

Young \& Knezek (1989) have shown that the ratio of molecular to atomic gas mass
in spiral galaxies increases from early type galaxies to late type galaxies by
a factor $\sim 10$. We have calculated the ratio
\begin{equation}
\frac{M_{\rm H_{2}}}{M_{\rm HI}}=\frac{\pi\,f_{\rm mol}\,\Sigma\,R^{2}}{\pi\,\Sigma\,R^{2}-
\pi\,f_{\rm mol}\,\Sigma\,R^{2}}=\frac{f_{\rm mol}}{1-f_{\rm mol}}\ .
\end{equation}
We used the following parameters: $Q=1$, $Re=50$, 
$\alpha=10^{7}$~yr\,M$_{\odot}$pc$^{-3}$, and $R$=10~kpc. 
The result can be seen in Fig.~\ref{fig:young_knezek}.
There is almost a factor 10 between the molecular to atomic gas ratio of a galaxy
with $v_{\rm rot}=50$~km\,s$^{-1}$ and a galaxy with $v_{\rm rot}=300$~km\,s$^{-1}$.
Of course, galaxies with small rotation velocities tend to be smaller,
but their H{\sc i} diameter should be larger, justifying $R$=const. as a first approach.
Furthermore, we expect that there is an influence of the molecule
formation time scale $\alpha$ on the ratio between molecular and atomic gas mass.
This time scale should depend on the galaxy's metallicity.
Thus the observed trend might be a mixture of (i) higher metallicities
(and therefore smaller $\alpha$) of early type spiral galaxies and (ii)
higher rotation velocities of early type spiral galaxies compared to
late type spirals, i.e. that early type galaxies are more massive.

\subsection{Star formation rate per unit mass}

The integrated star formation rate times the rotation period divided by the total 
mass should be comparable to the ratio of FIR luminosity to H band luminosity
$L_{\rm FIR}/L_{\rm H}$.
Using Eq.~\ref{eq:mdot} and $v_{\rm rot}^{2} \sim M_{\rm tot}G/R$, we obtain for our model
\begin{equation}
\frac{\dot{M}_{*}\Omega^{-1}}{M_{\rm tot}} \propto v_{\rm rot}^{-1}\ .
\end{equation}
This means that rapidly rotating galaxies have smaller $L_{\rm FIR}/L_{\rm H}$.
Devereux \& Hameed (1997) found indeed that galaxies of later type have smaller 
$L_{\rm FIR}/L_{\rm H}$. Since the FIR luminosity is assumed to be a measure
of recent star formation and the H band luminosity a measure of the total
mass, $(\dot{M}_{*}\Omega^{-1})/M_{\rm tot}$ decreases with increasing mass.
Moreover, Gavazzi et al. (1996) found a trend of smaller 
H$\alpha$ equivalent width for more luminous (H band) galaxies.
This also translates into a decreasing $(\dot{M}_{*}\Omega^{-1})/M_{\rm tot}$
with increasing galaxy mass. Thus, again, a part of this observed trend might be 
due to the fact that early spirals have higher rotation velocities than late type 
spirals.

\section{Summary}

We have analytically investigated the equilibrium state of a turbulent
clumpy gas disk. 
The disk consists of distinct self-gravitating clouds, which are be embedded
in a low density medium, and evolve in the fixed gravitational potential
of the galaxy. The disk scale height results from the balance between the
gravitational acceleration and the pressure due to the turbulent velocity
dispersion of the clouds. Gravitational cloud--cloud interactions in the
disk give rise to an effective viscosity and allows the transport of
angular momentum and mass in the gas disk. In our scenario turbulence
is assumed to be generated by instabilities involving self-gravitation 
and to be maintained
by the energy input, which is provided by differential rotation of the disk
and mass transfer to smaller galactic radii via cloud--cloud interactions.

A description of the turbulent viscosity is given, which is formally
equivalent to a prescription derived from the collisional Boltzman equation.
The vertical gravitational acceleration in the gas disk is either due to a
(i) dominating central mass, (ii) dominating stellar disk mass, or (iii)
the vertically self-gravitating gas disk itself. For all three cases
we derive analytical expressions for disk parameters (density, surface density,
velocity dispersion, disk height, driving wavelength, and viscosity)
as functions of
the Toomre parameter $Q$, the turbulent Reynolds number $Re$,
the mass accretion rate $\dot{M}$, the galactic radius $R$, and the angular 
velocity $\Omega$ of the gas.

The model does not include radiation. The ``thermostat'' equation of the
standard accretion disk model is replaced by an equation which assumes
that the energy flux generated by mass inflow and differential rotation
is entirely transported by turbulence to smaller scales where it is dissipated.
Whereas case (iii) corresponds to the a $Q \sim 1$ disk, case (i) and (ii)
require $Q > 1$.

The structure of the resulting clumpy gas disks allows us to derive  
global volume filling factors of self-gravitating clouds as a function of
$Q$ and $Re$. Along the same line analytical expressions for
the molecular fraction and the star formation rate are given.

On the basis of our analytical model we conclude that
\begin{enumerate}
\item
constant velocity dispersions in the gas as observed for spiral galaxies can
be reproduced by models with a (i) dominating central mass
and (iii) a vertically selfgravitating gas disk.
\item
In models (ii) with a dominating stellar disk the velocity dispersion
depends on the rotation velocity.
\item
The driving wavelength of the turbulence equals approximately the disk
scale height if the vertical acceleration is not dominated
by the stellar disk.
\item
The effective turbulent viscosity $\nu$ depends on the galactic radius.
For a disk in Keplerian rotation with a (i) dominating central mass 
$\nu \propto R^{\frac{3}{2}}$. The other two model classes (ii) and 
(iii) give $\nu \propto R$ for a constant velocity rotation curves.
\item
The molecular gas surface density and the star formation rate are proportional
to the square of the angular velocity 
$\Sigma_{\rm H_{2}},\ \dot{\Sigma_{*}} \propto \Omega^{2}$ in
self-gravitating gas disks.
\end{enumerate}

The model of a selfgravitating disk in vertical direction is applied to the
Galaxy and gives a good description of its gas disk. 
Furthermore, we investigate the r\^{o}le of the galaxy mass for the
morphological classification of spiral galaxies.

A comparison of our analytical model to spiral galaxies shows that
\begin{enumerate}
\item
the observed global star formation time scale implies a Reynolds number
$Re \sim 50$.
\item
our model is consistent with observations. That is the model reproduces
the physical parameters of the Galaxy as derived from observations.
\item
the derived mass accretion rate of the galaxy is 
$\dot{M} \sim 10^{-2}$~M$_{\odot}$\,yr$^{-1}$. It is thus
much smaller than the star formation rate $\dot{M}_{*}/\dot{M} \sim 100$. 
\item
the dependence of several physical parameters of spiral galaxies on morphological
type might be at least partly due to the mass--morphology relation, i.e.
galaxies of later types are more massive.
\end{enumerate}
In our model we assume that turbulence is generated by local instabilities 
due to selfgravitation. The energy input from accretion and differential 
rotation is high enough to maintain the ISM turbulence. Moreover,
the energy dissipation rate per unit mass due to the self-gravitation of the
gaseous disk in $z$ direction has the same order of magnitude as the
energy per unit time and unit mass transported to smaller scales by turbulence.
Thus, in terms of energy conservation, turbulence generated and
maintained by gravitation can account for the viscosity of spiral galaxies.
This does not exclude energy input in the turbulence through supernovae.
Nevertheless, our model where accretion in the galactic 
potential is sufficient to maintain ISM turbulence is fully consistent with 
observations.

\begin{acknowledgements}
We would like to thank the referee, K. Wada, for helping us to
improve this article considerably and W.J. Duschl for fruitful discussions.
\end{acknowledgements}

%---------------------------------------------------------------

\end{document}